\documentstyle[aps,prl]{revtex}
\begin{document}
\title{Time-like flows of energy-momentum and particle trajectories for the Klein-Gordon equation}
      \author{G. Horton, C Dewdney and A Nesteruk}
\address{Division of Physics,  University of Portsmouth. Portsmouth PO1 2DZ. England}
\maketitle
\begin{abstract}
The Klein-Gordon equation is interpreted in the de Broglie-Bohm manner as a single-particle relativistic quantum mechanical equation that defines unique time-like particle trajectories. The particle trajectories are determined by the conserved flow of the intrinsic energy density which can be derived from the specification of the Klein-Gordon energy-momentum tensor in an Einstein-Riemann space. The approach is illustrated by application to the simple single-particle phenomena associated with square potentials. 
\end{abstract}
\pacs{03.70,03.65}
\section{Introduction}
\label{introduction}
In a recent paper \cite{Dewdney96} we outlined a causal trajectory interpretation (in the de Broglie-Bohm sense) of the Klein-Gordon equation. We indicated how to proceed in order to retain the notion of particle trajectories, both for the single- and many-particle cases. The basic idea is to extend to the quantum context some results established in the context of general relativity for conserved flows of energy-momentum in classical scalar fields. In the classical context Edelen \cite{Edelen63}has shown that there is a natural definition of rest-energy flow and an associated conserved density which are determined respectively by the time-like eigenvectors and eigenvalues of the energy-momentum tensor of the scalar field itself in its local rest frame. In the context of a single-particle Klein-Gordon equation these time-like eigenvectors allow the definition of well-behaved particle trajectories. This approach overcomes some of the previous objections to the interpretation of relativistic scalar fields in terms of a particle ontology. 

In this paper we give further details of our approach and provide an illustration by showing how the simple quantum phenomena associated with single-particle scattering from static square potentials can consistently be interpreted in terms of well-defined individual relativistic boson trajectories. The single-particle picture is consistent and provides no requirement to have recourse to second quantization. (Particle creation can only be discussed in a many-particle approach.) We also discuss the Lorentz transformation properties of the trajectories.    

\section{Particle interpretations of the Klein-Gordon equation}
de Broglie first proposed a particle interpretation of the Klein-Gordon equation in the period 1926-1927\cite{deBroglie27}. de Broglie proceeded using the polar form of the scalar field,
\begin{equation}
\phi = R\exp(iS)
\end{equation}
to decompose the wave equation
\begin{equation}
\Box \phi=-m_0^2 \phi
\end{equation}
 into a continuity equation
\begin{equation}\label{eq:deBrogliec}
\partial^\mu (R^2\partial_\mu S) =0
\end{equation}
and a Hamilton-Jacobi equation
\begin{equation}
\partial_\mu S \partial^\mu S - \frac{\Box R}{R} =m_0^2
\end{equation}
This approach gives a positive definite Lorentz scalar probability density, $R^2$, but the problem with the use of $\partial_\mu S$ to define the flow lines is that $\partial_\mu S$ is not always a time-like four vector. Consequently the three-velocity, defined (in the absence of electromagnetic fields) by
\begin{equation}\label{eq:deBrogliev}
v=\frac{\nabla S}{-\frac{\partial S}{\partial t}}
\end{equation}
is not necessarily subluminal.
This fact was, of course, known to de Broglie \cite{deBroglie60} but he did not consider it a serious pathology. After all, no experimental consequences follow from the implied superluminal motion of the particles and the measurable predictions of the quantum theory are recovered and these do not contravene relativistic covariance. 

Another approach based on relativistic hydrodynamics has been developed (\cite{deBroglie26},\cite{Halbwachs60}, \cite{Vigier81}) 
but in this work also there is an implied assumption that $\partial_\mu S$ is a time-like four-vector. Because of the difficulty with space-like flows the particle interpretation of the Klein Gordon equation has not generally been accepted. Instead an interpretation based on well-defined, extended fields has been developed \cite{Bohm52},\cite{Bohm87},\cite{Lamm},\cite{Bohm93},\cite{Holland93}. 

It turns out however that at least for massive bosons there is a possibility to define time-like flows and hence particle trajectories that follow the flows and so one objection against a particle formulation of boson fields may be overcome. Our aim here is to explore the consequences of taking the idea of boson trajectories seriously. 

Dirac \cite{Dirac58} has shown that in relativistic hydrodynamics a Hamiltonian description can be given with reference to an arbitrary space-like hypersurface. The Hamiltonian density $H$ for any 
field splits into two pieces, each piece being proportional to a lapse or shift vector $(N,N_i)$. The two pieces have a simple physical significance as shown by Schutz \cite{Schutz71}, namely
\begin{eqnarray}
N\frac{\partial(H(-g)^{\frac{1}{2}})}{\partial N}&=&(-g)^{\frac{1}{2}}n^\mu 
T_{\mu\nu}n^\nu N \equiv \epsilon N\\
N_i\frac{\partial(H(-g)^{\frac{1}{2}})}{\partial N_i}&=&(-
g)^{\frac{1}{2}}g^{ij}n^\mu T_{\mu j} N_i \equiv P^i N_i
\end{eqnarray}
where $n^\mu = -N(^4g^{0\mu})$ is a unit normal to a space-like hypersurface, $(^4g^{\mu \nu})$ is 
the four dimensional metric and $g^{ij}$ is the three-dimensional 
metric. $\epsilon$ and $P_i$ are, respectively, the co-ordinate densities of energy and momentum measured by an observer at rest on the hypersurface.

One can define the flow lines for relativistic bosonic fields through the energy-momentum tensor $T^\mu_\nu$. In order to obtain a Lorentz invariant description one must form true 4-vectors to define the flow of energy momentum \cite{Holland93}. If one defines an arbitrary time-like vector $\eta^\nu$ then $T^\mu_\nu \eta^\nu$ is a true four-vector defining the flow of rest energy.\footnote{A justification for this approach can be found in \cite{Dirac58} and \cite{Schutz71}}. This flow, however, is dependent on the arbitrary choice of $\eta^\nu$ and is not therefore suitable, as it stands, if one wishes to ascribe beables to the field itself. 

Edelen \cite{Edelen63} has suggested that an intrinsic natural four-vector is provided by the matter field itself through the eigenvalue equation 
\begin{equation}
T^\mu_\nu W^\nu = \lambda W^\mu
\label{eigenvalue}
\end{equation}
Edelen shows that, given such a unique $W^\mu$, an intrinsic rest energy density $(\rho)$ exists and does not have to be introduced ad-hoc into the Einstein theory. The intrinsic energy density satisfies 
\begin{equation}
(\rho W^\mu)_{;\mu} = 0
\end{equation}
where $()_{;\mu}$ indicates the covariant derivative. 
Edelen establishes, given the unique $W^\mu$, that 
\begin{equation}
T^{\mu\nu} =  \lambda W^\mu W^\nu+\sigma^{\mu\nu}
\label{Tmunu}
\end{equation}
where $\sigma^{\mu\nu}$ is any symmetric tensor field with space-like support. Using also 
$(T^{\mu\nu})_{;\nu} = 0$ one gets
\begin{equation}
(\lambda W^\nu)_{;\nu} = \sigma^{\mu\nu}\epsilon_{\mu\nu}
\end{equation}
where $\epsilon_{\mu\nu}$ is the Born rate of strain tensor. One now looks for a path density $S$ such that its Lie derivative along $W^\mu$ equals 
$(-g)^{\frac{1}{2}}\sigma^{\mu\nu}\epsilon_{\mu\nu}$. It follows straightforwardly that
\begin{equation}
S=(\exp[ -\int^b_a \epsilon d\tau])\left (\int^{b}_{a}(\exp[\int^{\tau}_{a} \epsilon 
d\tau](-g)^{\frac{1}{2}}\sigma^{\mu\nu} \epsilon_{\mu\nu}d\tau \right )
\end{equation}
all integrations being performed along the $W^\mu$ and $\epsilon = 
\epsilon_{\mu\nu}g^{\mu\nu}$.

In the presence of electromagnetic fields we have
\begin{equation}
(T^{\mu\nu}_{matter})_{;\mu} = e j_{\mu}F^{\mu\nu}
\end{equation}
where $j_{\mu}$ is the charge current, $F^{\mu\nu}$ is the electromagnetic field tensor and $\partial_{\mu}F^{\mu\nu}$ is the Lorentz 4-force. The modified conservation relation now reads
\begin{equation}
(\lambda W^\nu)_{;\nu} = \sigma^{\mu\nu}\epsilon_{\mu\nu}+ e j_{\mu}F^{\mu\nu}W_{\nu}
\end{equation}
The path density is now defined to be
\[
S\left(s_0\right)=(\exp[ -\int^{s_0}_0 \epsilon d\tau])\left (\int^{b}_{a}(\exp[\int^{\tau}_{a} \epsilon d\tau](-g)^{\frac{1}{2}}\left(\sigma^{\mu\nu} \epsilon_{\mu\nu}+ j_{\mu}F^{\mu\nu}W_\nu\right) d\tau \right )
\]

$S$ will, in general, be path dependent and if the quantities in the integrand are continuous with no singularities can be set to zero on an initial space-like hypersurface. Therefore the initial value of $\rho$ is equal to $\lambda$ and $\lambda$ is a constant of the motion along the flow lines.

One now has a unique scalar density
\begin{equation}
\Phi = (-g)^{\frac{1}{2}}\lambda -S
\end{equation}
or, if one prefers, a unique $\rho$ given by 
\begin{equation}
\Phi=(-g)^{\frac{1}{2}}\rho
\end{equation}
One can say  that $\lambda (-g)^{\frac{1}{2}}W^\mu$ is a flux of rest energy density 
created in amounts equal to $\sigma^{\mu\nu}\epsilon_{\mu\nu}$ per unit geometrical 
volume. $\sigma^{\mu\nu}$ are the generalised stresses giving rise to the energy flow. $S$ 
then represents a path-dependent flux which combined with the flux of rest energy density 
gives a conserved flux $\Phi W^{\mu}$. $\Phi W^{\mu}$ can thus be thought of as an intrinsic energy density flux. 

Using this approach for the Klein-Gordon field circumvents a number of objections to the very notion of boson trajectories, either using the de Broglie approach or using the components of the energy-momentum tensor to define flow lines and densities \cite{Bohm87}. 

Other objections have been raised to single-particle formulations of relativistic quantum mechanics. For example there are a number of ways of setting up a Hilbert space for the Klein-Gordon equation \footnote{See for example Schweber \cite{Schweber}, Wald \cite{Wald94}. An example of a different approach is that of Gitman et al. \cite{Gitman90}.} but the scalar product utilised does not lead to a positive definite probability density. The position operator, also, does not have the usual form and it is not clear what measurement procedure would correspond to the measurement of position in the theory since the eigenfunctions are not delta functions and have an infinite range which suggests a violation of causality.

One way of proceeding has been to use the current $j^{\mu}$ which obeys a conservation relation
\begin{equation}
\int_V \partial_\mu j^\mu d^4x=\int_S j^\mu n_\mu d^3x
\end{equation}
where V is a four volume and S is an enclosing hypersurface with unit normal $n_\mu$. Since $j^\mu$ is not time-like everywhere, even in the case of a restriction to positive energy solutions, $j^\mu n_\mu$ does not provide a "probability" density analogous to the non-relativistic expression. In the case of trajectory theories one could use a scalar density $j^\mu n_\mu$ to give a density of crossings of the flow lines, given by $j_\mu$, across a hypersurface. The usual choice is to use equi-time surfaces giving a density of $j^0n_0$ (with $n_0=1$). A Lorentz transformation to a different frame of reference and equi-time hypersurface then leads to a density $\rho'$, given by
\begin{equation}
\rho'=j'_\mu n'^{\mu}=\rho
\end{equation}
One has however privileged one equi-time hypersurface (one does not choose $j_0$ in each frame). This defintion of $\rho$ does not, therefore, depend solely on the field but also, through $n^\mu$ on the choice of the space-like hypersurface. (D\"{u}rr et al. \cite{Durr} propose to treat the hypersurface as an additional dynamical variable with its own equation of motion.) One should note that the usual Klein-Gordon scalar product, defined via the conservation relation, is independent of the choice of surface $S$ and if one considers an infinitesimal tube along the direction of $j^\mu$ then putting $n_\mu=\frac{j_\mu}{\vert j_\mu \vert} $ one will have $ \vert j_\mu \vert d\sigma$ constant along the tube, where $d\sigma$ is the cross-sectional area normal to $j^\mu$. The fact that $j^\mu$ is not everywhere time-like, and hence the surfaces $\sigma$ not always space-like, means that $\vert j_\mu \vert $ is not interpretable as a probability density of finding a particle in a given position. The density of crossings of flow-lines is however given by $\vert j_\mu \vert$ which is a scalar quantity. In one frame of reference, at one point in space-time, there will be a rest frame in which $\rho=\vert j_0 \vert$ in the case of a time-like flow. Although this is not of use in the Klein-Gordon case it suggests that, in the Dirac case for example, one might choose the density of crossings as $\vert j_\mu \vert$ which will be independent of the hypersurface and Lorentz invariant. 

Although there may well be a preferential slicing of space-time \cite{Valentini} and equations of motion associated with that slicing, in view of the lack of a theory of such a preference, it seems better to suppose that the choice of space-like hypersurface for the preparation of a system in a given state and its subsequent measurement can be left open to the experimentalist (in thought, at least).

We therefore have proposed to take the eigenvalue $\lambda$ of the energy-momentum tensor $T^\mu_\nu$, corresponding to the unit time-like vector $W_\mu$ as the positive definite scalar density. In the Klein-Gordon case we have previously given the prescription for finding $W_\mu$ which always exists and solely depends on the structure of $T^\mu_\nu$. $\lambda$ being given by :-
\begin{equation}
\lambda=\frac{1}{2}\left[m_{0}^2\left|\phi\right|^2+\left|\partial_\mu\phi\partial^\mu\phi\right|\right]
\end{equation}
$\lambda$ is the rest energy at a point, in the frame of reference provided by $\left[ W^\mu\right]=[1,0,0,0]$ i.e. the locally stationary frame, and $\lambda \sqrt{-g}W^\mu$ is a flux of rest energy density $\rho\sqrt{-g}$ such that
\begin{equation}
\left(\rho W^\mu \right)_{;\mu}=0
\end{equation}
It is important to note that the flow lines given by $W^\mu$ are not geodesics since in the Klein-Gordon case in equation (\ref{Tmunu}) the $\sigma^{\mu\nu}$ is non-zero i.e. the Klein-Gordon field is not that of an incoherent fluid - there are additional state-dependent pressure terms of a purely quantum mechanical origin.
One sets $\rho$ equal to $\lambda$ on an arbitrary initial space-like hypersurface; $\rho\sqrt{-g}$ has been called the intrinsic energy density \cite{Edelen63}. The intrinsic energy density per unit geometrical volume is constant along the flow lines defined by $W^\mu$, i.e. $\lambda\sqrt{-g}da$ is constant along the flow lines and changes in $da$ are compensated by the $\sqrt{-g}$ factor, where
\begin{equation}
\frac{\sqrt{-g}}{\sqrt{\left(-g\right)_0}}=exp\left(+\int{\epsilon ds} \right)
\end{equation}
and $\epsilon=\epsilon^i_j$, $\epsilon_{ij}$ being the Born rate of strain tensor. 

$\lambda$ is, therefore, a constant of the motion \cite{Edelen63}. The advantage of using $\lambda$ as a density of crossings is that it is defined solely by the field and is a constant along the time-like flows; one may also consider arbitrary space-like hypersurfaces for preparation and measurement.

One must however note that $\lambda$ is a non-linear observable of the type proposed by Weinberg \cite{Weinberg} and Leifer \cite{Leifer} who give different accounts of a non-linear modification of quantum mechanics. Weinberg proposes a scheme that would enable one to measure such a non-linear observable. $\lambda$ is clearly not a conventional probability density. In justification of such a procedure one notes that in trajectory theories, such as those proposed by de Broglie and Bohm, there is no need to use the conventional apparatus of Hermitean operators and collapse to define the theory. One has a well-defined physical picture in which the particle follows a trajectory and once a measurement interaction is specified the outcome can be calculated for various initial particle co-ordinates. The uncontrollable nature of the initial co-ordinates in any given case limits the prediction to statistical statements only. One may therefore propose that the actual density of crossing of flow lines corresponds to that of the particles with no {\em necessity} to consider conventional measurement processes and no {\em need} to define a probability density yielding the statistical results if one were to make a measurement. One takes a consistent ontological position and accepts the contextualisation of results upon measurement.

The eigenvectors of $T_\mu^\nu$ are defined by (\ref{eigenvalue}). In the case 
of massive spin 0 and spin 1 fields unique time-like eigenvectors exist at each point of the Riemannian curved space-time. Transforming to Riemann normal coordinates gives $T^\mu_{\nu}$ in the usual flat space-time form (minimal coupling with no curvature contributions \cite{Fulling}. We discuss here only the scalar case.

Writing a solution $\phi$ of the Klein-Gordon equation as
\begin{equation}
\phi=\exp[P+iS]
\end{equation}
and setting $\partial_\mu P = P_\mu$ and $\partial_\mu S =S_\mu$
one gets \cite{Masden88},\cite{Wald94}
\begin{equation}
T_\nu^\mu = | \phi | ^2 [m_{0}^2-(P^\alpha P_\alpha+S^\alpha S_\alpha)] \delta^\mu_\nu +2 | \phi | 
^2 [ (P^\mu P_\nu+S^\mu S_\nu)]
\end{equation}
The second term can be written as $2|\phi|^2$ times
\begin{equation}
\left[\begin{array}{c}\vdots \\ P^\mu \\ \vdots \end{array} \right] [\ldots P_\mu \ldots] + 
\left[\begin{array}{c}\vdots \\ S^\mu \\ \vdots \end{array} \right] [\ldots S_\mu \ldots]
\end{equation}
the eigenvectors of $T^\mu_\nu$ must be a linear combination of $[P^\mu]$ 
and$[S^\mu]$ (or orthogonal to the term above). Some algebra shows that the two 
solutions are of the form (unnormalised):
\begin{eqnarray}
S^\mu + \exp^\theta P^\mu\\
S^\mu - \exp^{-\theta} P^\mu
\end{eqnarray}
where
\begin{equation}
\\sinh\theta = \frac{ P^\mu P_\mu-S^\mu S_\mu}{2P^\mu S_\mu}
\label{eq:theta}
\end{equation}

These two vectors are easily shown to be orthogonal four vectors using the expression for $\\sinh \theta$. One can conclude that one is time-like and the other space-like (or both null). If both eigenvectors were space-like then, in some frame of reference, their time components would both vanish giving $e^{\theta}=-e^{-\theta}=-\frac{S_{0}}{P_{0}}$ which is not possible with $\theta$ real as given by (\ref{eq:theta}).

In deriving $T^{\mu}_{\nu}$ one has assumed the usual metric with one time-like axis. As a result two further space-like vectors $a^\mu$ and $b^\mu$ can be derived such that 
\begin{eqnarray}
\nonumber
a^\mu\left(S_\mu+e^\theta P_{\mu}\right)&=&0\\
a^\mu\left(S_\mu-e^{-\theta P}_{\mu}\right)&=&0
\nonumber
\end{eqnarray}
and similarily for $b^{\mu}$, since the vectors $S_\mu+e^\theta P_{\mu}$ and $ S_\mu-e^{-\theta} P_{\mu}$ define a space-like two dimensional plane.
The overall eigenvalue $\lambda$ is given by
\begin{equation}
\label{eq:densityh}
\lambda=| \phi | ^2 \left[m_{0}^2\pm ([P^\mu P_\mu-S^\mu S_\mu]^2+4[P^\mu S_\mu 
]^2)^{\frac{1}{2}}\right]
\end{equation}
The three-velocity is given (in the absence of electromagnetic fields) by one of the two expressions
\begin{equation}
\label{eq:vh}
v=\frac{\nabla S \pm e^{\pm\theta}\nabla P}{-\left(\frac{\partial S}{\partial t}\pm e^{\pm\theta}\frac{\partial P}{\partial t}\right)}
\end{equation} 

In those regions where the de Broglie three-velocity is less than 1 then $\partial_\mu S$ is a time-like four vector in all frames of reference. A covariant condition can then be given for agreement between $v_{dB}$ and $v_e$.
Transforming to the rest frame at a point and rotating the space axis one gets
\[
e^\theta P_{1}'=-S_1',\: e^{-\theta} P_{0}'=S_0'
\]
or
\[
e^{-\theta} P_{1}'=S_1',\: e^{\theta} P_{0}'=-S_0'
\]
The second case can be dealt with by setting $\theta$ equal to $-\theta$ and $P_\mu'$ equal to $-P_\mu'$ in the first case, so one need only consider the first case in detail.
If one transforms back to a frame moving with three-velocity equal to $\tanh \alpha$ (at some point)one finds:
\begin{eqnarray}
\nonumber
S_1&=&-P_0\left(e^\theta+e^{-\theta}\right)\sinh \alpha \cosh \alpha-P_1\left(\cosh^2\alpha e^{-\theta}+\sinh^2 e^{\theta}\right)\\
S_0&=&P_0\left(\cosh^2\alpha e^\theta+\sinh^2 e^{-\theta}\right)+P_1\left(e^\theta+e^{-\theta}\right)\sinh \alpha \cosh \alpha
\nonumber
\end{eqnarray}
If the de Broglie three-velocity is less than one (at the considered point) one sees that $\theta$ must be sufficiently large and negative to get
\[
\frac{S_1}{-S_0}\rightarrow \tanh\alpha
\]
The two ways of calculating the velocity will agree where
\[
\left|\frac{S_1}{-S_0}\right|< 1
\]
with $\theta$ sufficiently large and negative (or positive for the other case).
In one dimension it can be shown that the $+$ signs give the time-like flows.
The flows of the field and hence the individual particle trajectories are described by a set of world-lines in space-time. Evidently they transform according to the appropriate Lorentz transformation to any other inertial frame. The world-lines of the particles are Lorentz invariant in the sense that the same set of events will be connected by the transformed world-line in all inertial frames. 

\section{Simple example: the square potential barrier}
In order to illustrate our physical model we compare the velocities and trajectories defined using the non-relativistic de Broglie-Bohm formulation, de Broglie's approach to the Klein Gordon equation and our definition in the context of single-particle scattering from a square potential barrier. The counter-intuitive behaviour which arises in this context is often cited as evidence of the inadequacy of single-particle interpretations of relativistic wave equations and of the need to proceed to second quantization, but on closer examination it can be seen that no paradoxical behaviour arises provided one determines the boundary conditions appropriately.  

Relativistically the potential may be either scalar or electrostatic in origin and we consider both cases here \footnote{The scalar potential was originally introduced to give a containment model for quarks (the so-called bag model) after it was found that an electrostatic delta-function barrier could not give containment.}. Using the relativistic relation
\begin{equation}
(\hbar\omega)^2=(\hbar k)^2+m_0^2
\end{equation}
(we take ${\hbar=c=1}$). In the regions of zero potential the momentum is given by 
\begin{equation}\label{eq:knonrel}
{\it k_1}=\pm \sqrt {{\omega}^{2}-{m_0}^{2}} \end{equation},
The momentum in the potential region depends on the type of potential. For the scalar case we have
\begin{equation}\label{eq:ksfsp}
{\it k_2}=\pm \sqrt {{\omega}^{2}-{(m_0+V)}^{2}}\end{equation}
whereas for the electrostatic case we have
\begin{equation}
\label{eq:ksfep}
{\it k_2}=\pm \sqrt {{(\omega-V)}^{2}-{m_0}^{2}}\end{equation} 
When dealing with an electrostatic potential step of semi-infinite extent one needs to take the positive square root in defining the momentum in the potential region in order to ensure that we have the correct physical boundary conditions. This choice avoids the appearance of waves travelling in from positive infinity when $E<V+m$ and hence the Klein paradox does not arise. \footnote{The original Dirac theory version of the paradox was effectively resolved in this manner \cite{bb}, see also the discussion in \cite{Greiner} p264. Pair production by strong fields is a reality, but evidently it can not be treated in a single-particle theory restricted to the single-particle sector of Fock space. Examples involving pair production can only be treated in many-particle theories.} We also take positive values for $\omega$, there is an exactly symmetrical set of solutions for negative $\omega$ but the two sets are disjoint. We shall consider in detail only the barrier case here. The electrostatic step behaviour cannot be obtained by extending the barrier width to infinity as one always has waves propagating in each direction when $E<V+m$. 

We define the potential barrier to occupy the region $0<x<a$ and to have magnitude {\it V} in a frame of reference that we label $\Sigma$.
In front of the barrier, for ${x<0}$ the wave function is \begin{equation}{\psi(x)=e^{i{\it k_1}\,x}}+R{e^{-i{\it k_1}\,x}}\end{equation} whereas in the region of the barrier we have
\begin{equation}{\psi(x)=G e^{i{\it k_2}\,x}}+H {e^{i{\it k_2}\,x}}\end{equation}
beyond the barrier
\begin{equation}{\psi(x)=J e^{i{\it k_1}\,x}}\end{equation}

Solving the boundary conditions at the barrier we find the standard expressions for the reflection and transmission ratios.
\begin{equation}
\left |T\right |^2=\frac{4 k_1^2 k_2^2}{\left(k_1^2+k_2^2\right)^2-\left(k_1^2-k_2^2\right)^2 cos^2(k_2a)}
\end{equation}
\begin{equation}
 \left |R\right |^2=\frac {\left(k_1^2-k_2^2\right)sin^2k_2a}{\left(k_1^2+k_2^2\right)^2-\left(k_1^2-k_2^2\right)^2cos^2(k_2a)}
\end{equation}
The different cases studied here are distinguished by the prevailing relationship between $k$ and $\omega$. In the scalar case $k_2$ becomes imaginary when ${V > \omega-m}$ (and the so-called Klein paradox does not arise for scalar potential steps), whereas in the electrostatic case $k_2$ is imaginary only in the interval ${\omega-m < V < \omega+m}$, for which there is exponential decay in the barrier. As the barrier height is increased beyond $\omega+m$, $k_2$ becomes real and one has a transmitted and a reflected wave in the barrier region and there is no confinement.

In the usual approach the physical meaning of the component wave functions is derived from their associated currents, one speaks in terms of an incoming, transmitted and reflected current. But the wave function in front of the barrier is, of course, a pure superposition state and so strictly there is just the overall current associated with this pure state. Where a plane-wave description is used, the incoming and reflected wave functions overlap over the whole of space $(x<0)$ and so, in the usual approach, for the purpose of interpretation the wave function is tacitly considered to be a mixture; only then can the physical situation be described in terms of separate incident and reflected currents. The reflected current is given by $-{\it k_1}\left |R\right |^{2}$ and the transmitted current by ${\Re({\it k_2}}\left |T\right |^{2} )$. In the pure state the current is always positive or zero. As the reflection coefficient increases the current decreases, vanishing as the reflection coefficient approaches unity. 

For the scalar potential the variation of the transmission and reflection coefficients for a given energy and over a range of barrier heights and widths is similar to that familiar from the non-relativistic case. {\footnote We set $m=1$ for the calculations presented in this section.} As shown in Fig. 1. the transmission decays to zero as the barrier height increases to $V=\omega-m$ and the wave is completely reflected. For an electrostatic barrier the situation is somewhat counter intuitive. For an incident momentum of 0.95 (in our arbitrary units), Fig. 2. shows the behaviour of the transmission coefficient (for the electrostatic case) as the magnitude and width of the barrier are varied. At first, as expected, the transmission decreases as the barrier potential height or width increases and this continues until the region for which ${V=\omega \pm m}$ throughout which there is an exponential decay of the wavefunction in the barrier. As the magnitude of the barrier potential increases still further, transmission once more increases right up to ${V=2\omega}$ where full transmission occurs. Thereafter transmission oscillates with increasing barrier height and width (where in the scalar or non-relativistic case the transmission is zero).

In the non relativistic case the continuity equation for the conserved density ${\rho=|\psi|^2}$ is
\begin{equation}
\frac{\partial \rho}{\partial t}={\nabla . \left(\rho \vec v \right)}
\end{equation}
and this suggests that the velocity be defined by
\begin{equation}
\vec v_S=\frac{\vec j}{\rho}=={\it Im}({\frac {{\frac {\partial }{\partial x}}\psi(x,t)}{\psi(x,t)}}
\end{equation}
which for the region in front of the barrier yields
\begin{equation}
v_S={\frac {{\it k_1}\,\left (1-{R}^{2}\right )}{2\,R\left (\cos({\it 2 k_1}
\,x)\right )+1+{R}^{2}}}
\end{equation}
At the minima of the denominator the velocity is given by
\begin{equation}
v_S={\frac {{\it k_1}\,\left (1+{R}\right )}{\left (1-{R}\right)}}
\end{equation}
Evidently for the non-relativistic case the velocity has no upper bound and where the density is very small the velocity will be very large. The density will develop nodes as ${R \to 1}$. This case was solved exactly by Takabyasi \cite{Takabayasi}. Typical trajectories for the incident wave packet case can also be seen in \cite{Dewdney82}.

De Broglie's expression for the velocity in the Klein Gordon case is given in equation (\ref{eq:deBrogliev}). For the particular case of the barrier this yields 
\begin{equation}
v_{dB}=\frac{1}{(\omega-V)}v_{S}
\end{equation}
with ${k_1}$ in the expression for $v_S$ now given by (\ref {eq:ksfep}), for the region in front of the step. In the limit for which ${k_1}$ is small the de Broglie velocity approaches $v_S$. Again there is no upper bound on the velocities defined in this way. As the reflection coefficient approaches unity the regions around the minima for which the velocity is superluminal shrink, but the magnitude of the velocity in these regions increases rapidly. The regions shrink to a point as the reflection coefficient approaches unity.

The situation is markedly different if one uses the time-like eigenvector of the energy-momentum tensor to define velocities and its eigen-value to define the density as described above in equations (\ref{eq:vh}) and (\ref{eq:densityh}). In this particular one-dimensional case, for the definite energy eigenfunctions, the expression for the velocity can be written
\begin{equation}
v_e=\frac{\nabla S + e^{\theta} \nabla P}{\omega-V}
\end{equation}
This velocity is a weighted combination of the de Broglie velocity and what has been referred to elsewhere as the osmotic velocity. The combination always has a magnitude less than 1.

For this simple case where the wave function in front of the barrier is a superposition of two counter-propagating waves with complex weights, it can be shown that the eigenvalue of the energy-momentum tensor $\lambda$ and $\left|\phi\right|^2$ are in phase and related at the maxima and minima by 
\begin{equation}
\lambda=2k^2\left|R\right|+\left(1\pm\left|R\right|\right)^2
\end{equation}
In this case then, even for complete reflection when $R=1$, although $\left|\phi\right|^2$ will go to zero at the minima, $\lambda$ remains finite everywhere. 
\footnote{This is similar to the Dirac case where the density can not have nodes\cite{Dewdney92}.} 
If one wants to calculate $\rho$ at later times then he piecewise constant potential introduces some complications since the delta-function forces at the beginning and end of the barrier introduce discontinuities in the path density S and one will have to integrate along the paths. We have
\begin{equation}
j_{\mu}F^{\mu\nu}W_\nu=ej^\mu\left(\frac{\partial A_\nu}{\partial x^\mu}-\frac{\partial A_\mu}{\partial x^\nu}\right)W^{\nu}
\end{equation}
For the one-dimensional, time-independent electrostatic barrier this becomes
\begin{equation}
j^{\mu}F_{\mu\nu}W^\nu=e\left(j^3W^0-j^0W^3\right)\frac{\partial A_0}{\partial x^3}
\end{equation}
which has a singularity at the start and end of the barrier, hence there will be a step discontinuity in S at the beginning and end of the barrier.

For the purposes of illustration we calculate the various densities, trajectories and velocities defined above, in the lab frame in which the barrier is stationary, for a variety of cases.
 
For a momentum of $k_1=0.1$ (in our arbitrary units in the $\Sigma$ frame) the particle has a velocity of one tenth of the speed of the light. If the reflection coefficient is also low then the three expressions for the velocity are approximately equal, as are the densities and the motion is time-like in all cases.  The situation is very different if we consider a large reflection coefficient, say $R=0.99$. With the relativistic energy given by $E=\sqrt{1.01}$ the Schr\"odinger  velocity and the de Broglie velocity are similar, the relativistic scaling is just $\sqrt{1.01}$. Both expressions give superluminal motion in the interference minima. The eigen-vector velocity remains sub-luminal showing only a small oscillation. As the reflection coefficient approaches unity the de Broglie velocity increases without limit at the minima of $\left|\phi\right|^2$, whereas the energy-momentum velocity approaches zero as $\lambda$ approaches a minima. A graph comparing the de Broglie and the energy-momentum velocities and densities is given in Fig. 3. It is clear that even in this "non-relativistic" limit the relativistic corrections to the velocity are important in maintaining a sub-luminal velocity in the interference minima. That the low-energy limit of the relativistic de Broglie-Bohm velocity and the non-relativistic de Broglie-Bohm velocity differ considerably in interference situations was first noted in \cite{Dewdney92}.The de Broglie trajectories associated with this situation are given in Fig. 4., and the eigen-vector trajectories in Fig. 5. The relativistic trajectories at low energy differ considerably from those calculated using the non-relativistic (or even the de Broglie) formulation. Several authors have suggested using non-relativistic Bohm trajectories to calculate low-energy barrier-tunnelling times, however we see that the calculation of tunnelling-times based on the low-energy relativistic trajectories will in general give rather different predictions.

At the relativistic momentum $k=0.95$ the differences are more marked for all values of the reflection coefficient. Fig. 6 shows a comparison of the velocities and densities defined in the de Broglie theory and in ours for the case of a reflection coefficient of $0.7$ when $E<V-m$ (the normal region). Fig. 7. shows the associated de Broglie trajectories and Fig. 8. the associated energy-momentum trajectories. Note that the velocities differ appreciably in the region of the interference minima. When $V-m<E<V+m$ the reflection coefficient is unity and the velocities are zero. When $V>E+m$ we enter the Klein paradox region, Fig. 9. shows the de Broglie and eigen-vector velocities and densities for a reflection coefficient of $0.7$ in the anomolous transmission region. Fig. 10. shows the de Broglie trajectories associated with the velocities of Fig. 9. Note that the motion is always in the positive direction. Fig. 11. shows the eigen-vector trajectories associated with the velocities of Fig. 9.
 
\section{Lorentz covariance}
In the theory we have given here, based on the eigenvalues and eigenvectors of the energy momentum tensor in the local rest frame of the flow, the density is a Lorentz scalar and the flows of energy-momentum are defined by four-vectors and are always time-like. The Klein-Gordon wave function, which determines the density and the flow evolves unitarily in a covariant manner and so the problems associated with wave-packet collapse in a relativistic theory do not arise in the same way (there is no wave packet collapse). If the particle position (or the entire world line) is taken as the {\it beable} or the {\it element of reality} then these are clearly Lorentz covariant. For a given space-time wave function $\Phi\left(x\right)=\Phi'\left(x'\right)$, there is a unique world line through a given space-time event. The set of space-time points linked by a world line is invariant, but their co-ordinate description in other frames is, of course, different. So the situation here is no different than that for a classical particle world-line. We labour this point as doubts have been raised concerning the Lorentz covariance of trajectories as defined in relativistic de Broglie-Bohm theories. 

To expand on this point further, consider a single-particle system prepared in the laboratory or $\Sigma$ frame on a space-like hypersurface (given by the actual experimental details), typically this will be an equal-time hypersurface, although it need not be so. It is important to remember here that all preparations must be referred to some extended space-like hypersurface and not to a single point or to a time-like hypersurface. From this initial specification on the arbitrary space-like hypersurface the evolution of the wave function can be calculated and the space-time energy-momentum flow lines are then determined using the prescription given above. The calculation of the wave function on any other arbitrary space-like hypersurface, different from the initial surface, would have to be carried out using a type of Tomanaga-Schwinger approach, evolving the initial wave function to the chosen hypersurface. The flow lines remain the same, irrespective of the choice of hypersurface, although their co-ordinatisation using alternative frames of reference will be different. There is an important distinction to be made between the arbitrary choice of space-like surface and an inertial given frame.

Consider now the same physical situation as it is described in another inertial frame, $\Sigma'$, in motion with respect to $\Sigma$. In this frame the system is not described as having been prepared on an equal-time hypersurface, the initial hypersurface in $\Sigma'$ is just the Lorentz transformed hypersurface defined in $\Sigma$. The relationship between the flow-lines in $\Sigma$ and $\Sigma'$ is simply that of the usual passive re-coordinatisation of world lines which follows on the application of the appropriate Lorentz transformation.
There is no preferred frame in this approach, although there is, as a matter of contingent fact, a particular space-like hypersurface on which the system is actually prepared. 

It is perhaps not surprising that Lorentz covariance follows naturally in the single-particle case where non-local correlations cannot arise. We consider the many-particle case in a further paper. In the approach that we have described here measurements must be treated as particular physical interactions described by introducing the appropriate interaction terms in the Hamiltonian and the appropriate additional apparatus variables. There is no wave packet collapse and the wave equation will still yield a covariant process evolving unitarily into different channels each associated with a different outcome of the measurement {\cite{Bohm93}. In a further paper we will consider the Klein-Gordon equation for a fixed number of particles greater than one and its interpretation in terms of particle trajectories. This will allow consideration of Lorentz covariance and nonlocality in a relativistic boson particle trajectory de Broglie-Bohm theory.

\newpage
{\bf Figure Captions}
\begin{enumerate}
\item{FIG. 1. Variation of the transmission coefficient with magnitude $V$ and width $a$ of a scalar potential barrier for an incident energy of 1.38.}
\item{FIG. 2. Variation of the transmission coefficient with magnitude $V$ and width $a$ of an electrostatic potential barrier for an incident energy of 1.38.}
\item{FIG. 3. Schr\"odinger, de Broglie (dotted line) and eigenvector (solid line) velocities and densities for $k=0.1$,$a=12$, $V=0.0306$ and $R=0.99$, the Schr\"odinger and de Broglie velocities are not distinguishable, the eigenvector velocities are always less than one.}
\item{FIG. 4.De Broglie trajectories and their direction-field for $k=0.1$,$ a=12$, $V=0.0306$ and $R=0.99$ in the region of an interference minima.}
\item{FIG. 5. Energy-momentum flow trajectories and their direction-field for $k=0.1$,$ a=12$, $V=0.0306$ and $R=0.99$ in the region of an interference minima.}
\item{FIG. 6.De Broglie (dotted line)and energy-momentum (solid line)velocities and densities for $k=0.95$,$a=12$,$V=0.36$ and $R=0.7$.}
\item{FIG. 7. De Broglie trajectories and their direction field for $k=0.95$,$a=12$,$V=0.36$ and $R=0.7$.}
\item{FIG. 8. Eigenvector trajectories and their direction field for $k=0.95$,$a=12$,$V=0.36$ and $R=0.7$.}
\item{FIG. 9. de Broglie (dotted line)and energy-momentum (solid line)velocities and densities for $k=0.95$,$a=12$,$V=4.47$ and $R=0.7$, in the Klein-region.}
\item{FIG. 10. De Broglie trajectories and their direction field for $k=0.95$,$a=12$,$V=4.47$ and $R=0.7$, in the Klein region.}
\item{FIG. 11. Eigen-vector trajectories and their direction field for $k=0.95$,$a=12$,$V=4.47$ and $R=0.7$, in the Klein region.}
\end{enumerate}

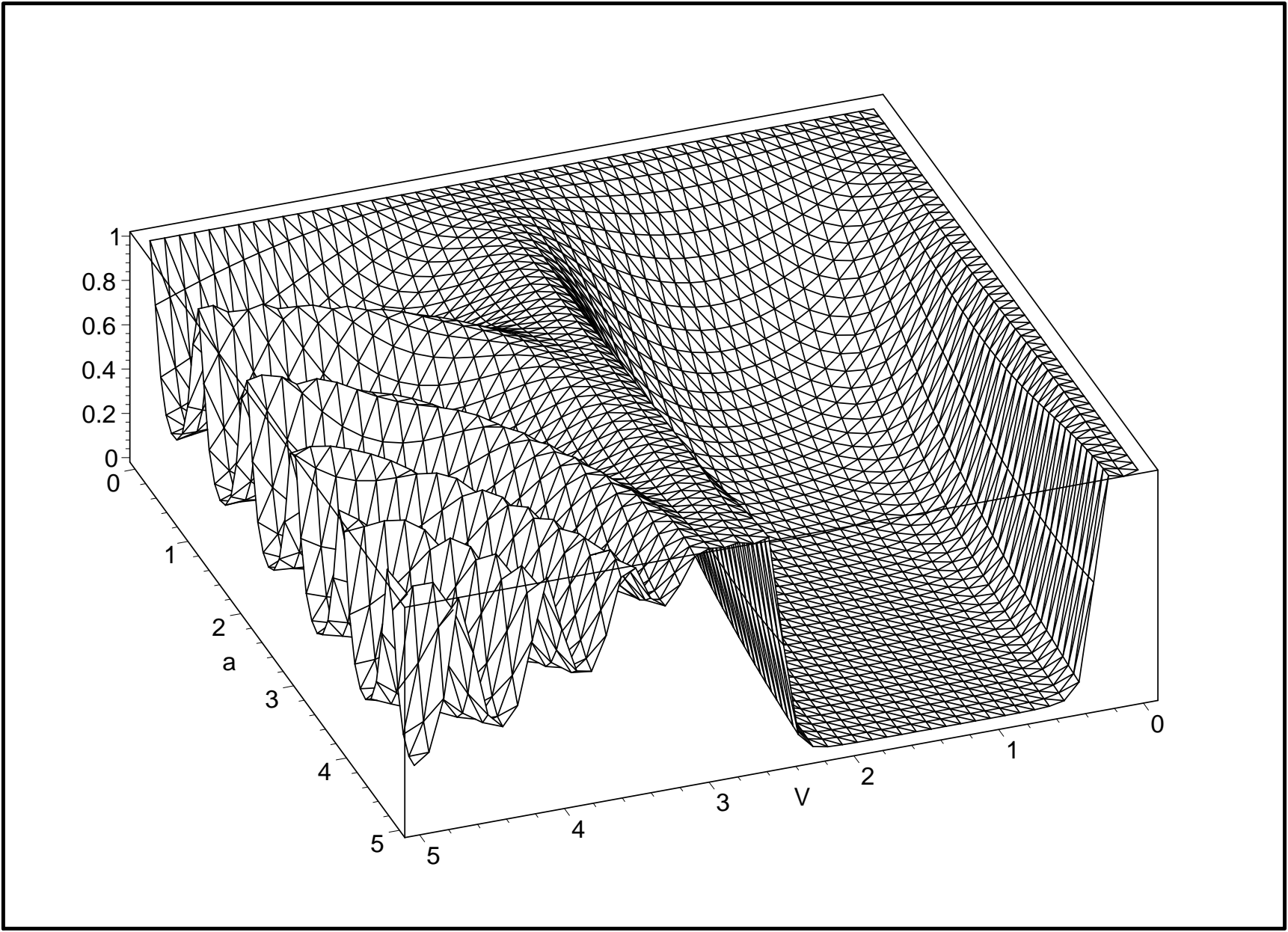
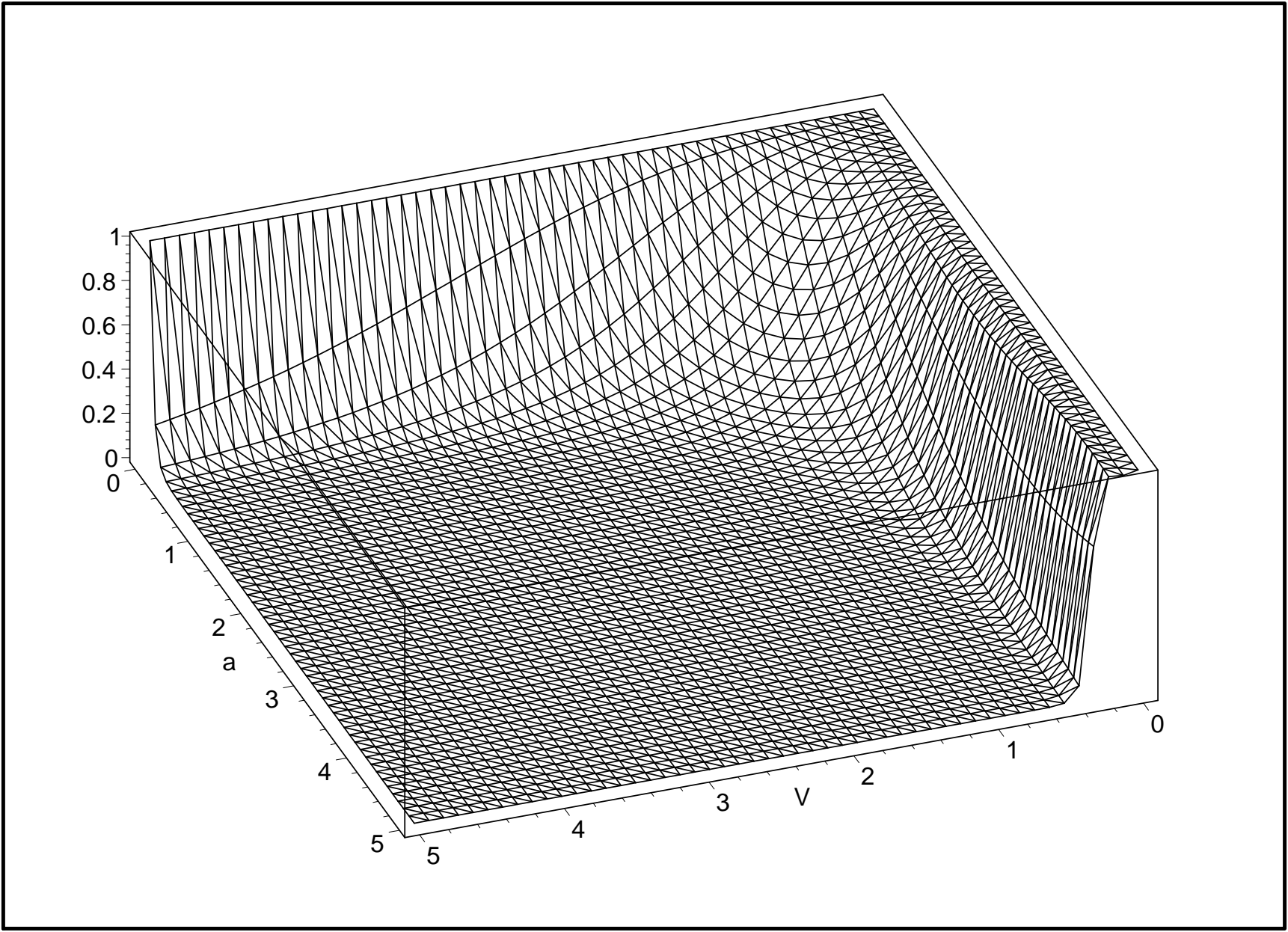
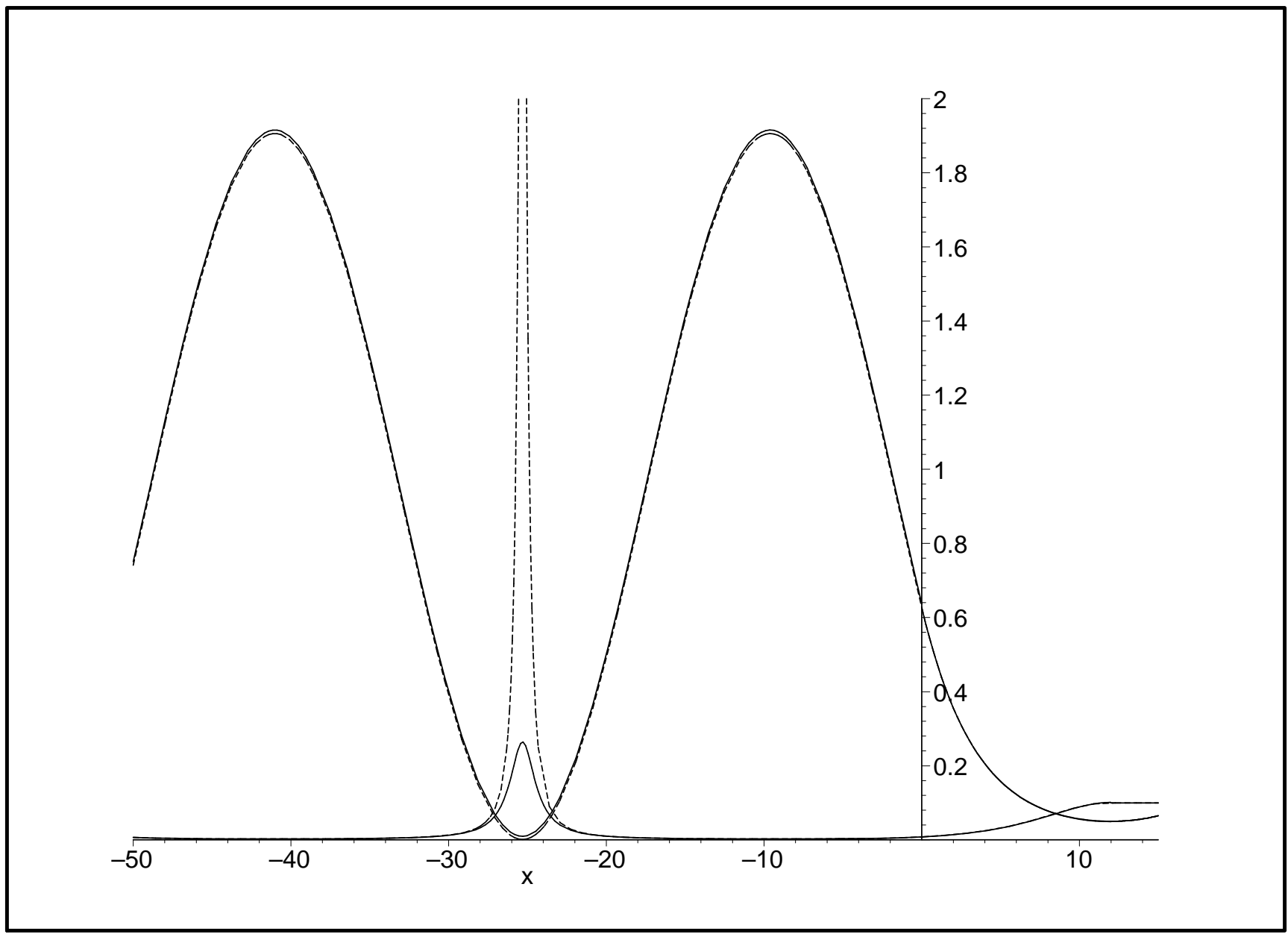
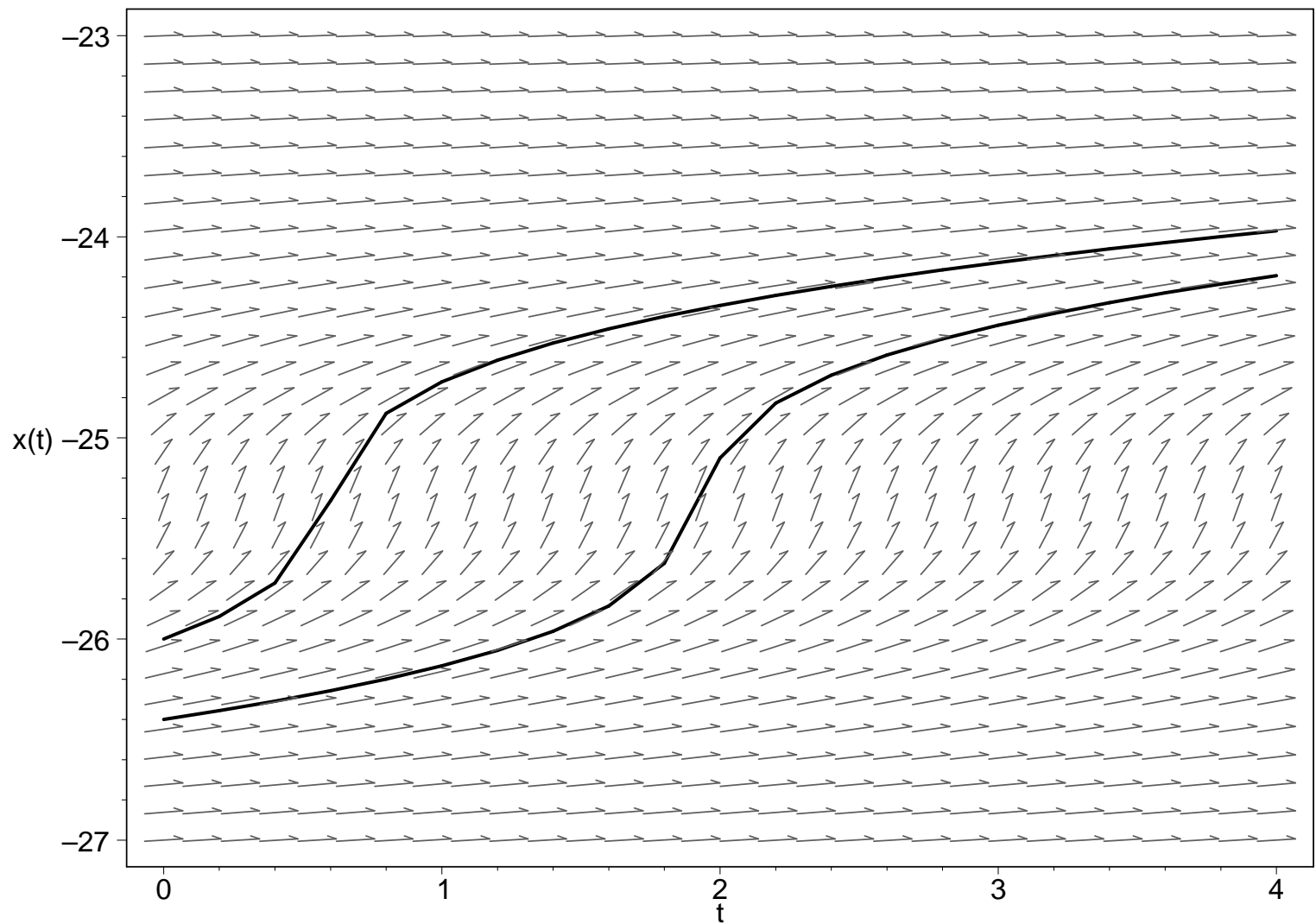
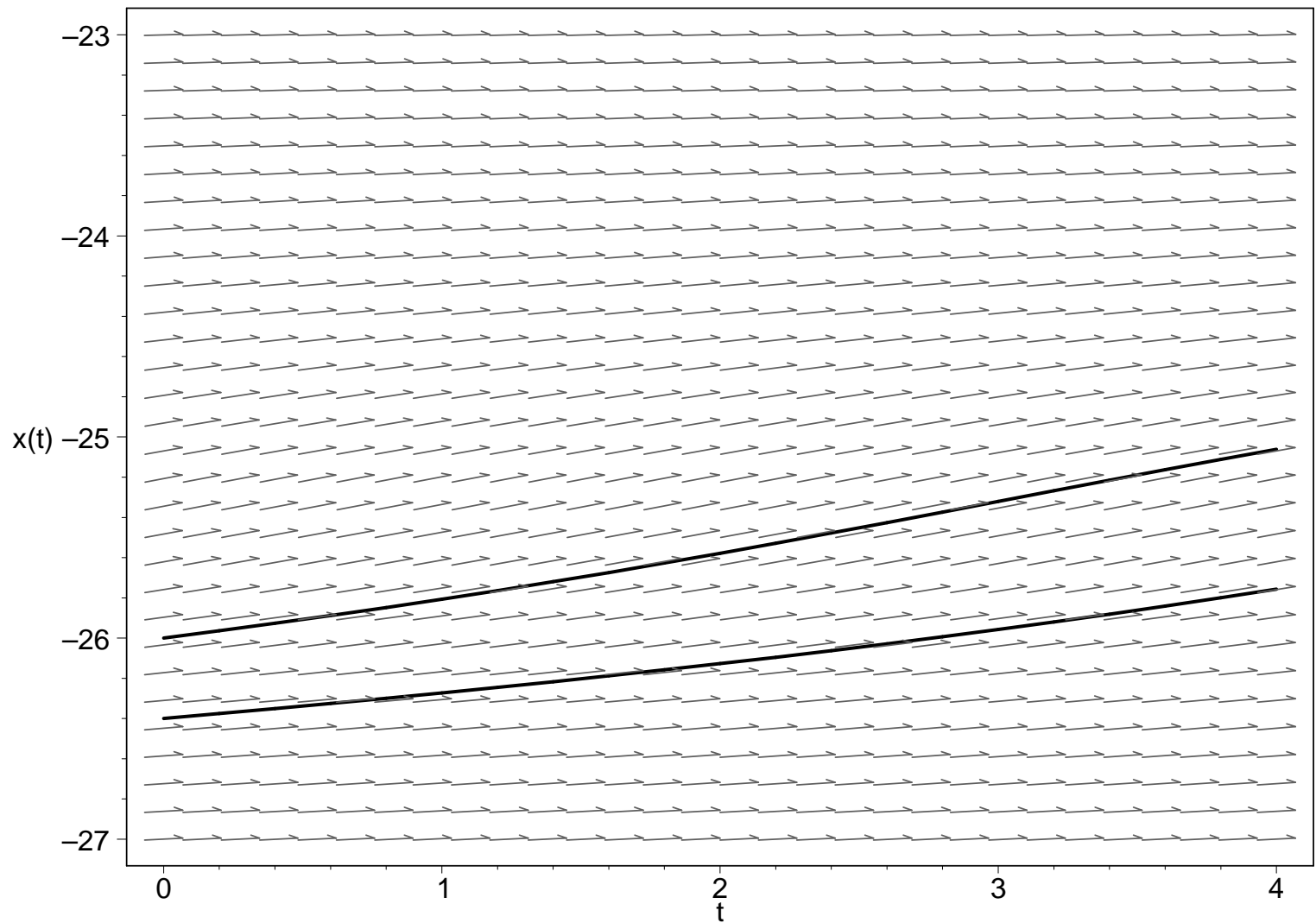
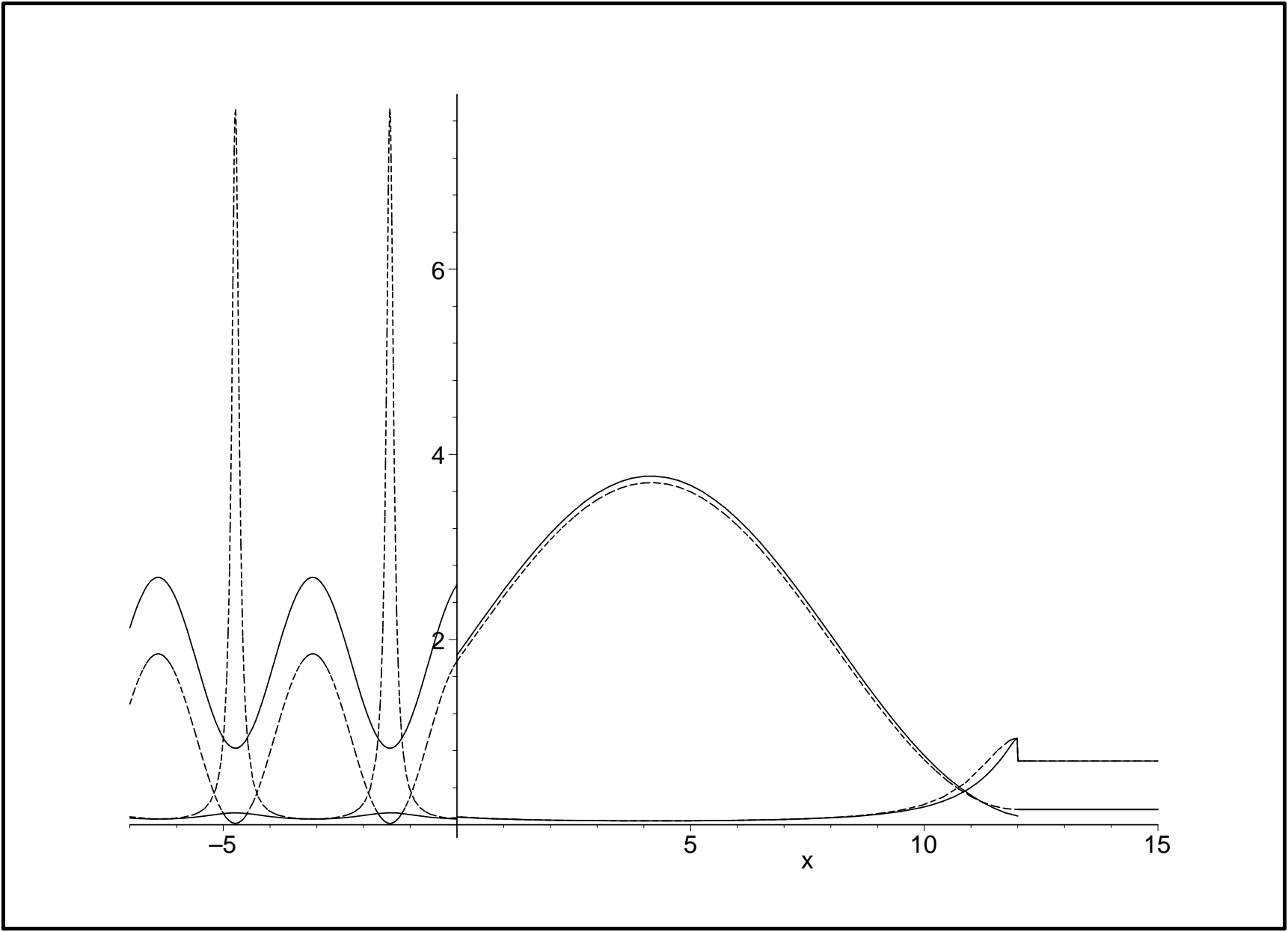
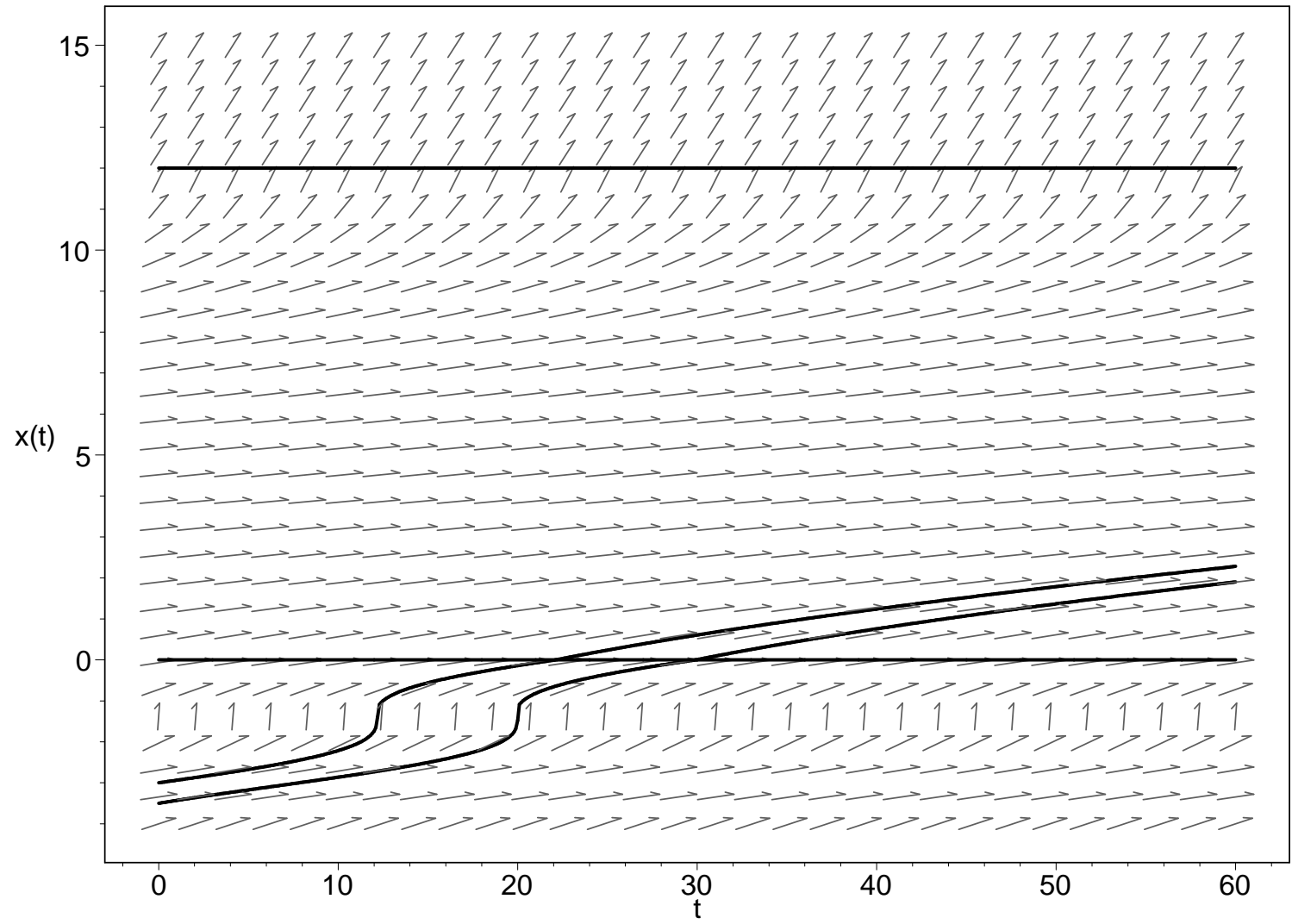
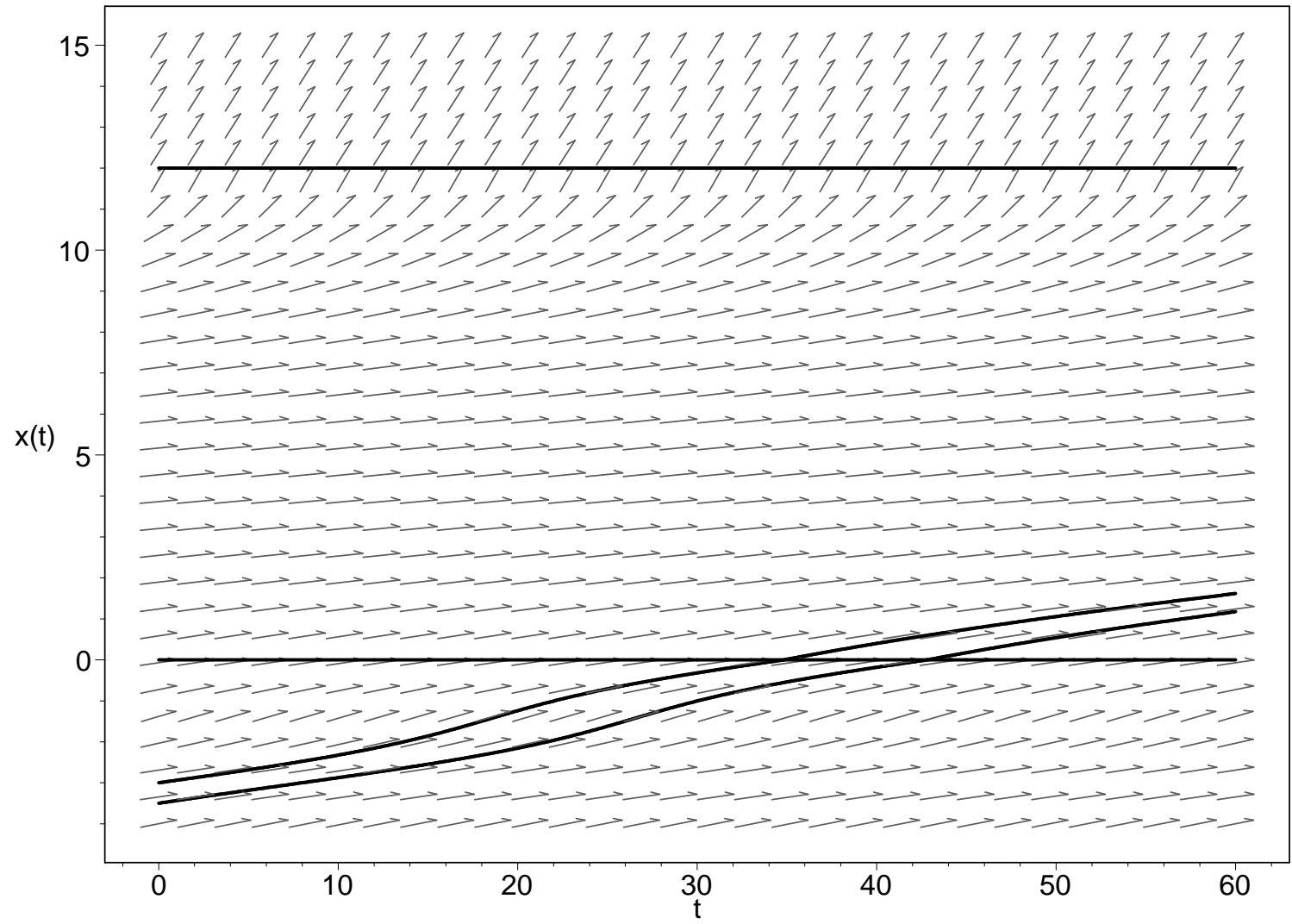
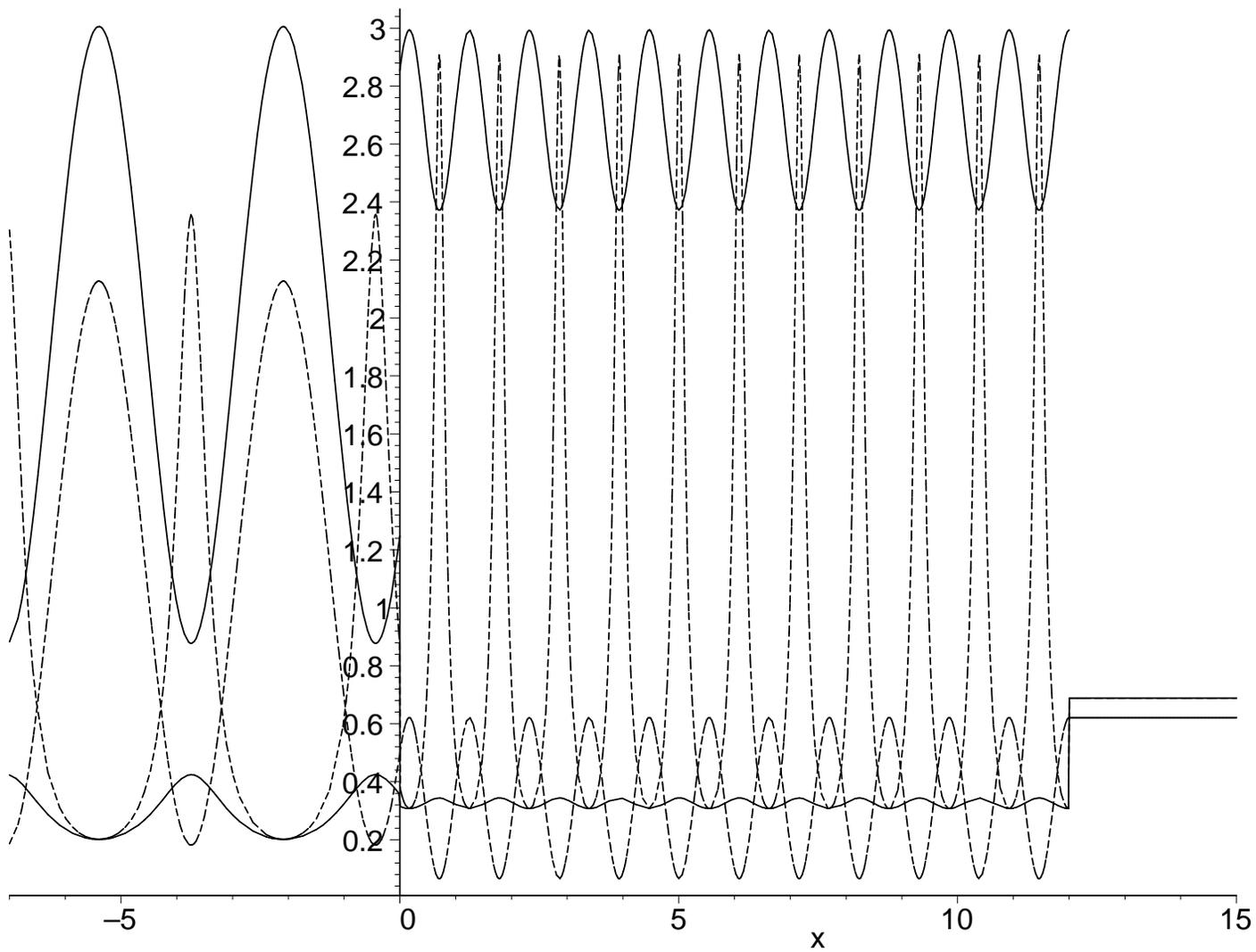
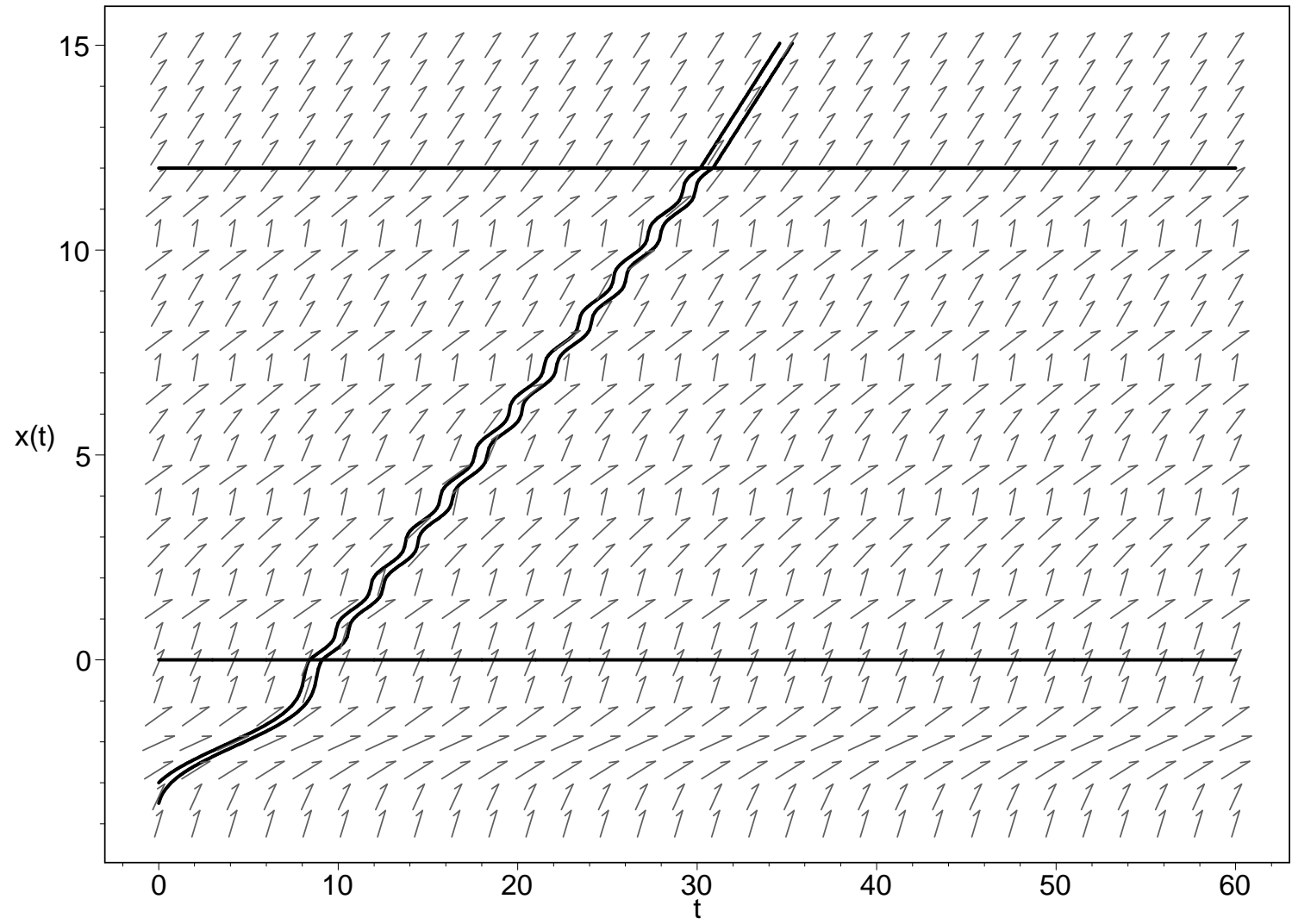
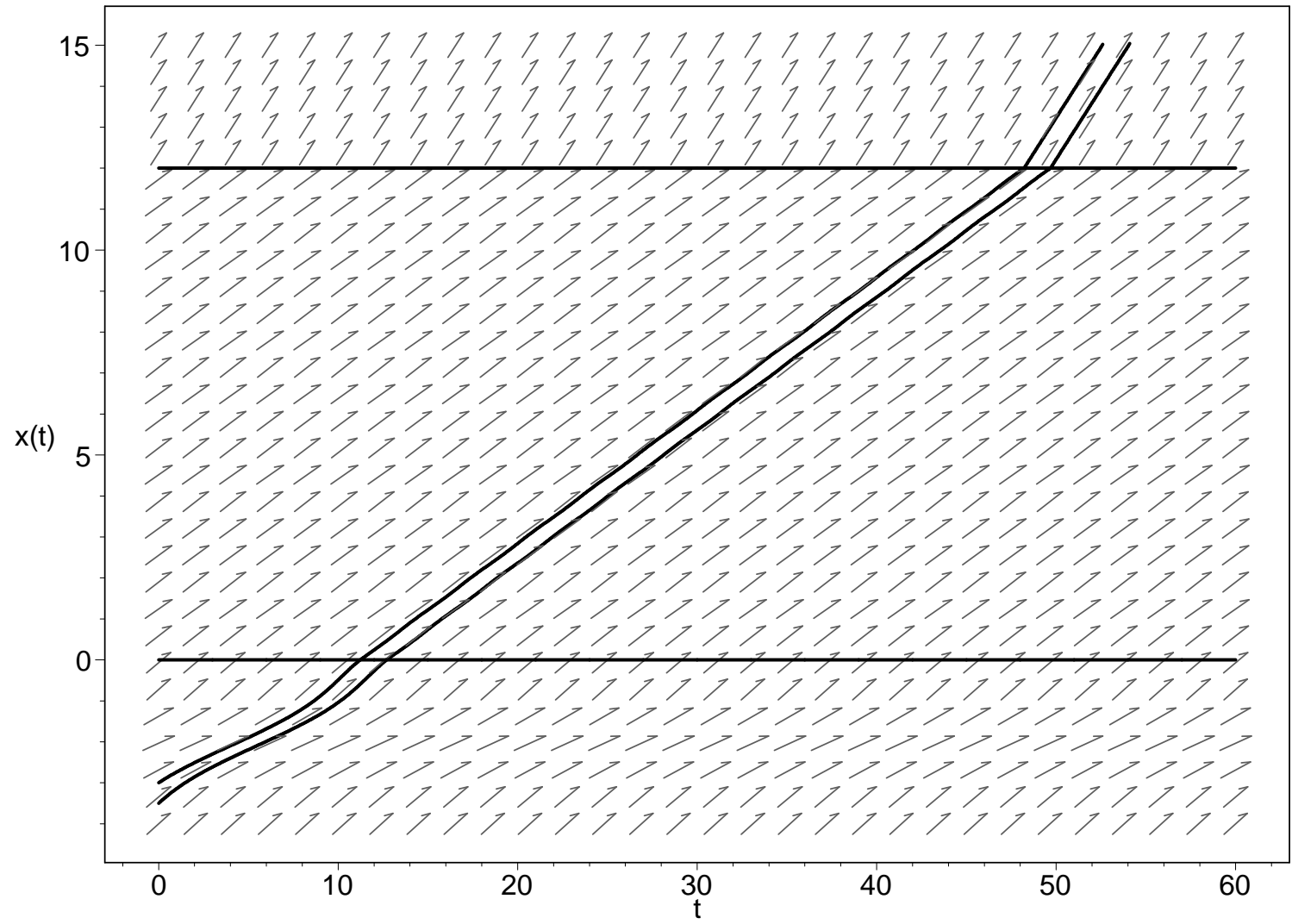
\end{document}